\documentclass[aps,prl,preprint,tightenlines,superscriptaddress]{revtex4}

\usepackage{graphicx} 
\usepackage{dcolumn}  
\usepackage{epsfig,amsmath,amssymb,euscript}
\graphicspath{{ps}}

\renewcommand{\arraystretch}{1.1}

\newcommand{\Fig}[1]{Figure~\ref{#1}}

\newcommand{\Tab}[1]{Table~\ref{#1}}

\RequirePackage{xspace}

\def\gev    {\ensuremath{\,{\rm GeV}}}

\def\gevcd  {\ensuremath{\,{\rm GeV}/\mathit{c}^2}}
\def\gevdcd {\ensuremath{\,{\rm GeV}^2/\mathit{c}^2}}

\def\mbSF    {\ensuremath{m_b(\rm SF)}}
\def\mupidSF {\ensuremath{\mu_{\pi}^2(\rm SF)}}

\def\Bxulnu  {\ensuremath{B \to X_u \ell \nu}\xspace}

\def\bsg     {\ensuremath{B \to X_s \gamma}}
\def\Vubs    {\ensuremath{|V_{ub}|}\xspace}

\begin{document}

  \title{ \quad\\[0.5cm] Determination of the $b$-quark leading shape
  function parameters in the shape function scheme using the Belle
  \bsg\ photon energy spectrum}

\affiliation{J. Stefan Institute, Ljubljana}
\affiliation{High Energy Accelerator Research Organization (KEK),
  Tsukuba}

  \author{Ilija~Bizjak}\affiliation{J. Stefan Institute, Ljubljana} 
  \author{Antonio~Limosani}\affiliation{High Energy Accelerator Research Organization (KEK), Tsukuba} 
  \author{Tadao~Nozaki}\affiliation{High Energy Accelerator Research Organization (KEK), Tsukuba} 

\collaboration{of the Belle Collaboration for the Heavy Flavor
  Averaging Group}

\noaffiliation

\begin{abstract}
We determine the $b$-quark shape function parameters in the shape
function scheme, \mbSF\ and \mupidSF\, using the Belle \bsg\ photon
energy spectrum.  We assume three models for the form of the shape
function; exponential, gaussian and hyperbolic.
\end{abstract}

\maketitle
\section{Introduction}
\label{intro}

One of the best ways to determine the Cabbibo-Kobayashi-Maskawa(CKM)
matrix element \Vubs\ is to measure the inclusive charmless
semileptonic (\Bxulnu) partial branching fraction of the $B$
meson. Its theoretical prediction relies on an accurate description of
non-perturbative bound-state effects of a $b$ quark in the $B$
meson. The effects are encoded in $B$-meson shape functions, which are
not theoretically calculable and have to be determined
experimentally. The leading order shape function can be determined
using the photon energy spectrum in \bsg\ decays, since up to the
leading order of $1/m_b$ in the Heavy Quark Expansion it is described
with the same leading order shape function as the one used for the prediction
of the \Bxulnu\ decays. Such shape function determination was first
performed by CLEO on their \bsg\ data~\cite{Gibbons}, and was then repeated in
Ref.~\cite{limonoz} for the Belle \bsg\ data, where the theory
proposed by Kagan and Neubert~\cite{KN} was used to describe the
photon energy spectrum.

Recently, Bosch, Lange, Neubert and Paz presented theoretical expressions
for the triple differential \Bxulnu\ decay rates and for the
\bsg\ photon spectrum, which incorporate all known contributions and
smoothly interpolate between the "shape-function region" of large
hadronic energy and small invariant mass, and the "OPE region" (in
which all hadronic kinematical variables scale with
$M_B$)~\cite{BLNP,generator}. The differential rate is given in terms
of a leading shape function and a subleading shape function in the
shape function scheme~\cite{BLNP,generator}, where in this scheme the
leading shape function is expressed with the parameters \mbSF\ and
\mupidSF.

In this paper we report on the determination of the shape function
parameters by fitting the Belle \bsg\ data with the predicted photon
energy spectrum~\cite{generator}, where the default subleading shape
function model from Ref.~\cite{generator} is used. It allows for a
precise determination of $|V_{ub}|$ as shown in Ref.~\cite{endpoint}
and Ref.~\cite{fullrec}.

Apart from using an updated and more complete theoretical description
of the \bsg\ decays, this analysis differs from the one in
Ref.~\cite{limonoz} by using a different set of shape function model
parametrizations, namely exponential, gaussian and hyperbolic. It also
includes subleading shape function effects that were absent in the
former analysis.

\section{Procedure}

We used a method based on that devised by the CLEO
Collaboration~\cite{Anderson}. We fit Monte Carlo (MC) simulated
spectra to the raw data photon energy spectrum. ``Raw'' refers to the
spectra that are obtained after the application of the \bsg\ analysis
cuts. The use of ``raw'' spectra correctly accounts for Lorentz
boost from the $B$ rest frame to the center of mass system, energy
resolution effects and avoids unfolding. The method is as follows:
\begin{enumerate}
\item Assume a shape function model.
\item Simulate the photon energy spectrum for a certain set of
  parameters; (\mbSF, \mupidSF).
\item Perform a $\chi^2$ fit of the simulated spectrum to the data
where only the normalization of the simulated spectrum is floated and
keep the resultant $\chi^2$ value.
\item Repeat steps 2-3 for different sets of parameters to construct
a two dimensional grid with each point having a $\chi^2$.
\item Find the minimum  $\chi^2$ on the grid and all points on the grid
that are one unit of  $\chi^2$ above the minimum.
\item Repeat steps 1-5 for a different shape function model.
\end{enumerate}


\subsection{Shape function models}
Three shape function forms suggested in the literature are employed;
exponential, gaussian and hyperbolic~\cite{generator}. Their
functional forms are described in \Tab{tab:sfforms}: they are a
function of $\hat\omega$ and are parameterized by two parameters:
$\Lambda$ and $b$. Example shape function forms are plotted in Fig. 1.
\begin{table}[hbpt!]
  \begin{center}{
      \renewcommand{\arraystretch}{1.8}
      \begin{tabular}{cc} \hline\hline
        Shape Function & Form \\ \hline
        exponential & $F^{\rm (exp)}(\hat\omega;\Lambda,b) =
        \frac{N^{\rm (exp)}}{\Lambda}\left(\frac{\hat\omega}{\Lambda}\right)^{b-1}
        \exp\left( -d_{\rm exp} \frac{\hat\omega}{\Lambda}\right)$ \\
        gaussian & $F^{\rm (gauss)}(\hat\omega;\Lambda,b) =
        \frac{N^{\rm (gauss)}}{\Lambda}\left(\frac{\hat\omega}{\Lambda}\right)^{b-1}
        \exp \left(-d_{\rm gauss}\frac{\hat\omega^2}{\Lambda^2} \right)$ \\
        hyperbolic & $F^{\rm (hyp)}(\hat\omega;\Lambda,b) =
        \frac{N^{\rm (hyp)}}{\Lambda}\left(\frac{\hat\omega}{\Lambda}\right)^{b-1}
        \cosh^{-1} \left(d_{\rm hyp}\frac{\hat\omega}{\Lambda}\right)$ \\  \hline
        \multicolumn{2}{l}{where the constants are:}\\
        \multicolumn{2}{c}{
	$\begin{aligned}
	N^{\rm (exp)} &= \frac{d_{\rm (exp)}^b}{\Gamma(b)} \,, 
    	& d_{\rm (exp)} &= b \,, \\
   	N^{\rm (gauss)} &= \frac{2\,d_{\rm (gauss)}^{b/2}}{\Gamma(b/2)} \,, 
    	& d_{\rm (gauss)} &= \left(\frac{\Gamma\left(\frac{1+b}{2} \right)}%
    	{\Gamma\left(\frac{b}{2} \right)}\right)^2 \,, \\
   	N^{\rm (hyp)} &= \frac{[4\,d_{\rm (hyp)}]^b}{2\,\Gamma(b) 
    	\left[ \zeta(b,\frac14) - \zeta(b,\frac34) \right]} \,,
    	& d_{\rm (hyp)} &= \frac{b}{4} \, 
    	\frac{\zeta\left(1+b,\frac14\right) - \zeta\left(1+b,\frac34\right)}%
    	{\zeta\left(b,\frac14\right) - \zeta\left(b,\frac34\right)} \,,
	\end{aligned}$}\\
	\multicolumn{2}{c}{$\zeta(b,a)=\sum_{k=0}^\infty (k+a)^{-b}$ is the generalized
        Riemann zeta function}\\	
	\hline\hline
	\end{tabular}\\
      }
    \end{center}
    \caption[Shape function models]{The three models used for the shape function forms\label{tab:sfforms}}
  \end{table}

The parameters $\Lambda$ and $b$ are related to the HQET parameters
$\bar{\Lambda}$ and $\mu_\pi^2$ by analytical expressions Eq.46 and
Eq.47 in Ref.~\cite{generator} for exponential and gaussian models,
respectively, while for the hyperbolic model the corresponding HQET
parameters have to be calculated numerically. The shape function
parameters \mbSF\ and \mupidSF\ are obtained from the HQET parameters
$\bar{\Lambda}$ and $\mu_{\pi}^2$ using the relations in Eq.41 of
Ref.~\cite{generator}, where the reference scale of $1.5 \gev$ is used.
\begin{figure}[hbpt!]
\begin{center}
\begin{tabular}{cc}
\includegraphics[width=0.50\columnwidth]{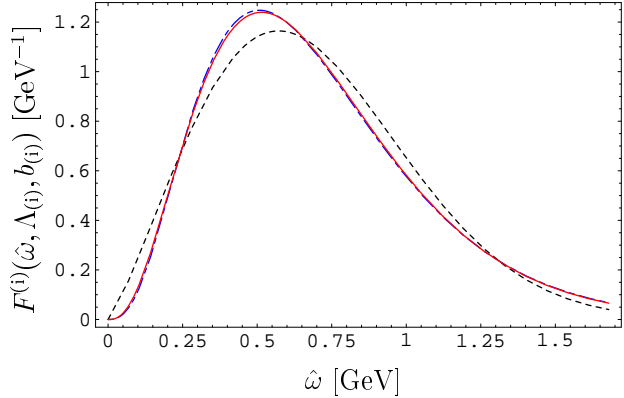} 
\end{tabular}
\end{center}
\caption[Shape function forms]{
Three different models with linear onset for small $\hat\omega$. We
show $F^{\rm (exp)}$ (solid), $F^{\rm (gauss)}$ (dashed), and $F^{\rm
(hyp)}$ (dash-dotted), for parameters that correspond to \mbSF=
4.63\gevcd\ and \mupidSF= 0.2 \gevdcd~\cite{generator}.\label{fig:sfcurves}}
\end{figure}


\subsection{Monte Carlo simulated photon energy spectrum}

We generate \bsg\ MC events according to the prescription in
Ref.~\cite{generator} for each set of the shape function parameter
values. The generated events are simulated for the detector
performance using the Belle detector simulation program and the \bsg\
analysis cuts are applied to the MC events to obtain the raw photon
energy spectrum in the $\Upsilon(4S)$ rest frame~\cite{Koppenburg}.


\subsection{Fitting the spectrum}

For a given set of shape function parameters, a $\chi^2$ fit of the MC
simulated photon spectrum to the raw data spectrum is performed in the
interval $1.8 <E^*_\gamma < 2.8 \gev$~\footnote{The $*$
denotes the center of mass frame or equivalently the $\Upsilon(4S)$
rest frame.}.
Although in the Ref.~\cite{limonoz} the fitting was performed in the
interval between $1.5 \gev$ and $2.8 \gev$, the data below $1.8 \gev$
are not used in the present analysis since the tails of the models we
use below $1.8 \gev$ are not modelled accurately~\cite{lange}.

\begin{figure}
\begin{center}
\includegraphics[width=0.50\columnwidth]{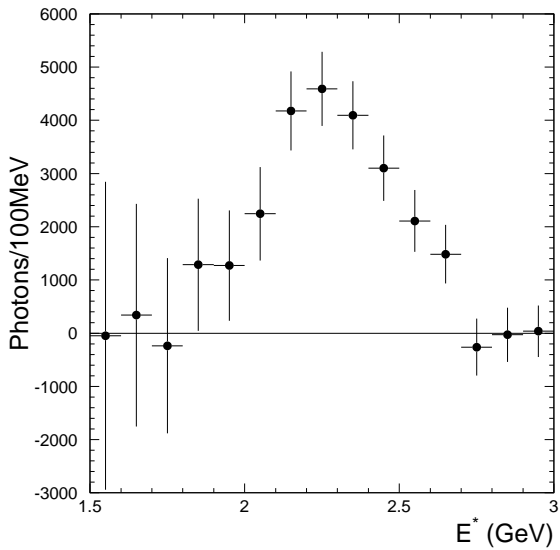} \\
\end{center}
\caption{ Raw $B \rightarrow X_s \gamma$ photon energy spectra in the
$\Upsilon(4S)$ frame as acquired from data. The errors include both
statistical and systematic errors.  Raw refers to spectra as measured
after the application of Belle $B \rightarrow X_s \gamma$ analysis
cuts. \label{fig:bsgraw}}
\end{figure}

The normalization parameter is floated in the fit. The
raw spectrum is plotted in \Fig{fig:bsgraw}, the errors include both
statistical and systematic errors. The latter are dominated by the
estimation of the $B\overline{B}$ background and are 100\% correlated.
Therefore the covariance matrix is constructed as
\begin{equation}
V_{ij} =
\sigma^{\mathrm{stat}}_{d_i}\sigma^{\mathrm{stat}}_{d_j}\delta_{ij}
+   \sigma^{\mathrm{sys}}_{d_i}\sigma^{\mathrm{sys}}_{d_j}  
\end{equation}
where $i,j=1,2,\ldots,10$ denote the bin number, and $\sigma_d$ is the
error in the data. Then the $\chi^2$ used in the fitting is given by
\begin{equation}
\chi^2 = \sum_{ij} (d_i-f_i) (V^{-1})_{ij} (d_j-f_j),
\end{equation}
where $(V^{-1})_{ij}$ denotes the $ij^{th}$ element of the inverted
covariance matrix.
The $\chi^2$ value of the fit is used to determine a map of $\chi^2$ 
as a function of the shape function  parameters.
\subsection{The best fit and $\Delta \chi^2$ contour}

The best fit parameters are associated to the minimum chi-squared
case, $\chi^2_\mathrm{min}$. The error ``ellipse'' is defined as the
contour which satisfies $\Delta\chi^2 \equiv
\chi^2-\chi^2_\mathrm{min}=1$. The contours are found to be well
approximated by the function\cite{Fac},
\begin{equation}\label{eqn:ellipse}
\Delta \chi^2(\mbSF,\mupidSF)=
\left( \frac{\mupidSF + a (\mathit{\mbSF})^2 + b }{c} \right )^2 +
\left( \frac{(\mathit{\mbSF})^2 + d}{e} \right )^2. 
\end{equation}
The parameters $a$, $b$, $c$, $d$, and $e$ are determined by fitting
the function to the parameter points that lie on the contour.


\section{Results}
\begin{figure}[hbpt!]
\begin{center}
  \begin{tabular}{ccc}
    \includegraphics[width=0.33\columnwidth]{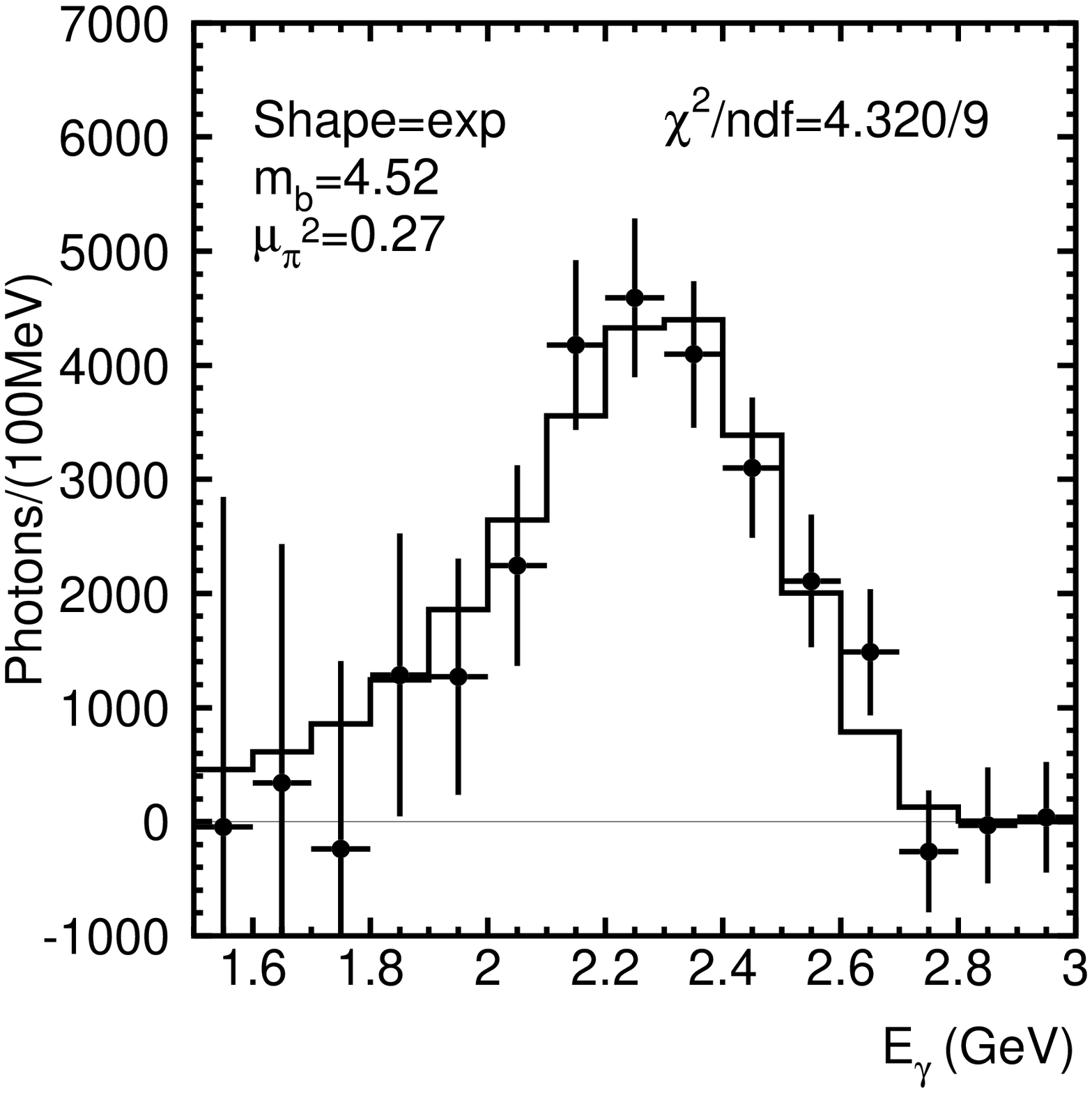} & 
    \includegraphics[width=0.33\columnwidth]{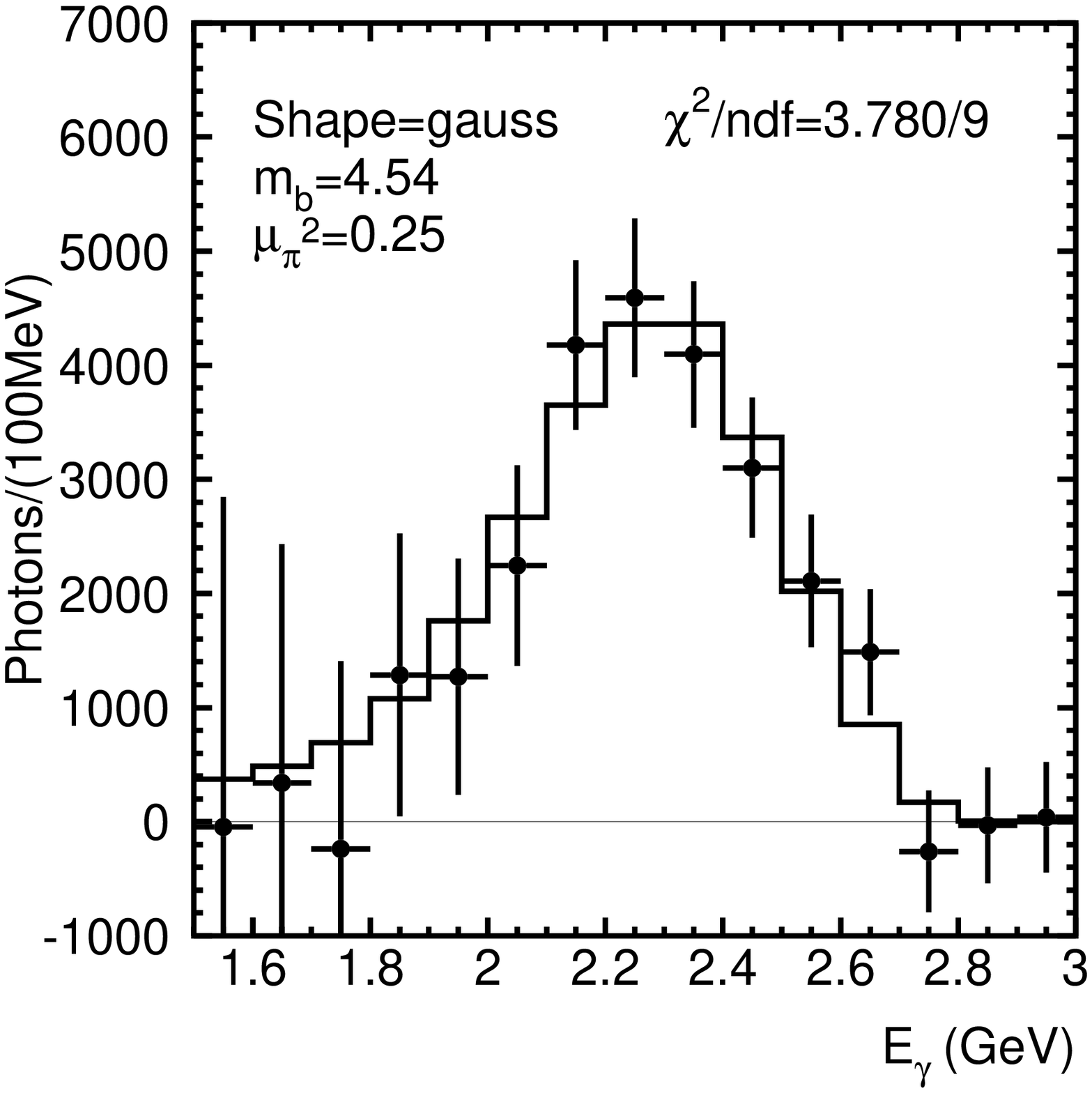} &
    \includegraphics[width=0.33\columnwidth]{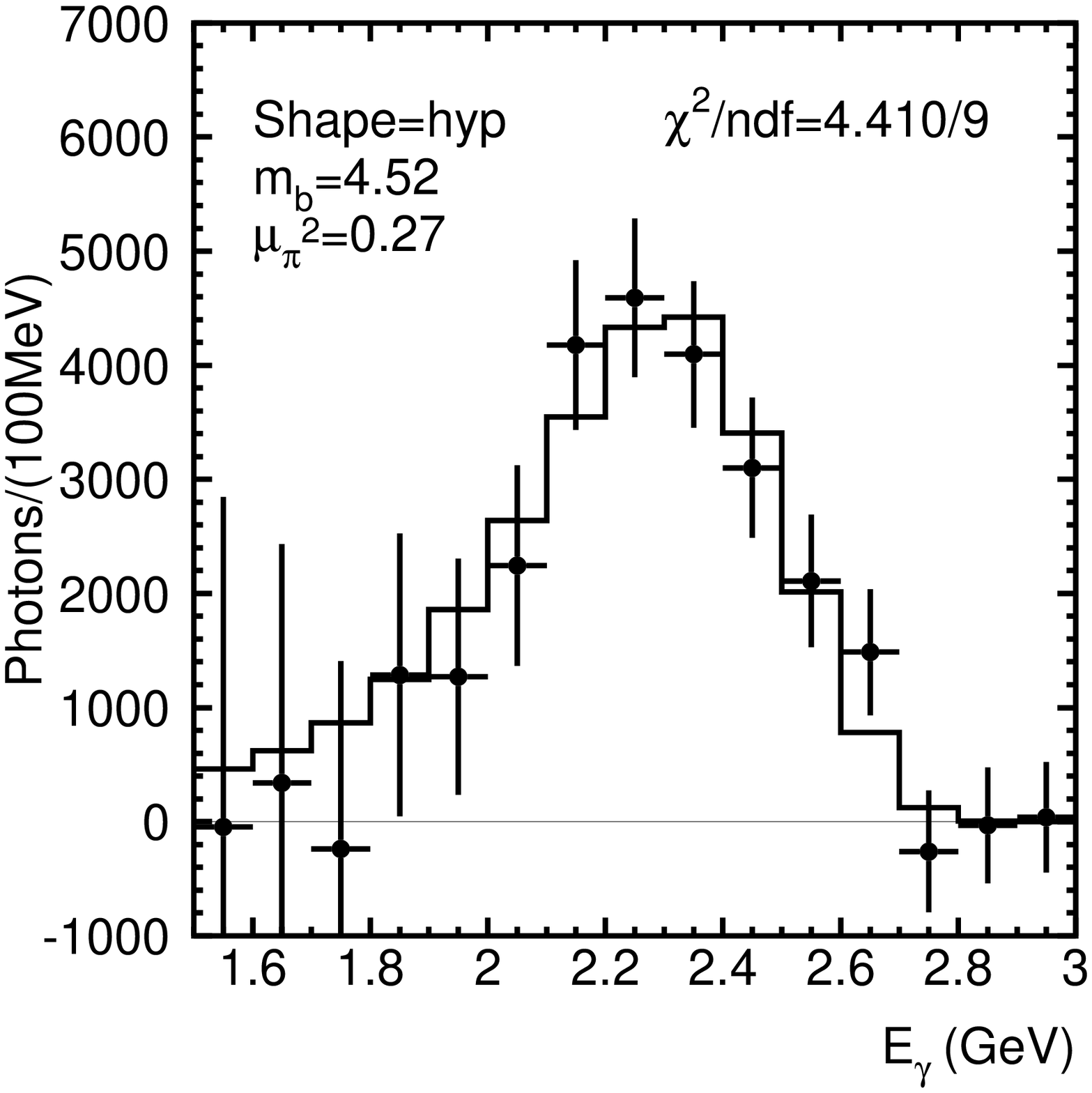} \\
    \includegraphics[width=0.33\columnwidth]{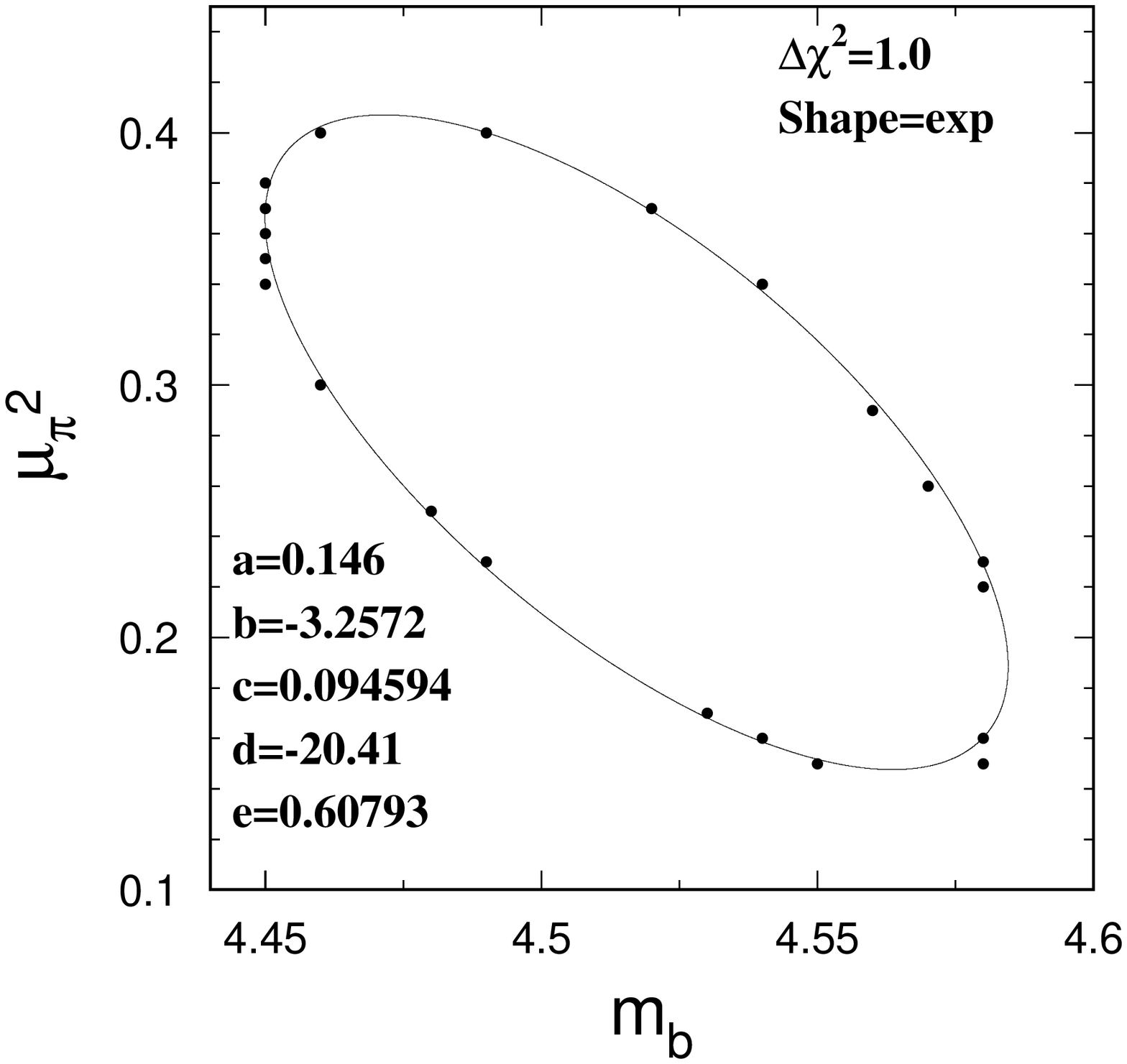} & 
    \includegraphics[width=0.33\columnwidth]{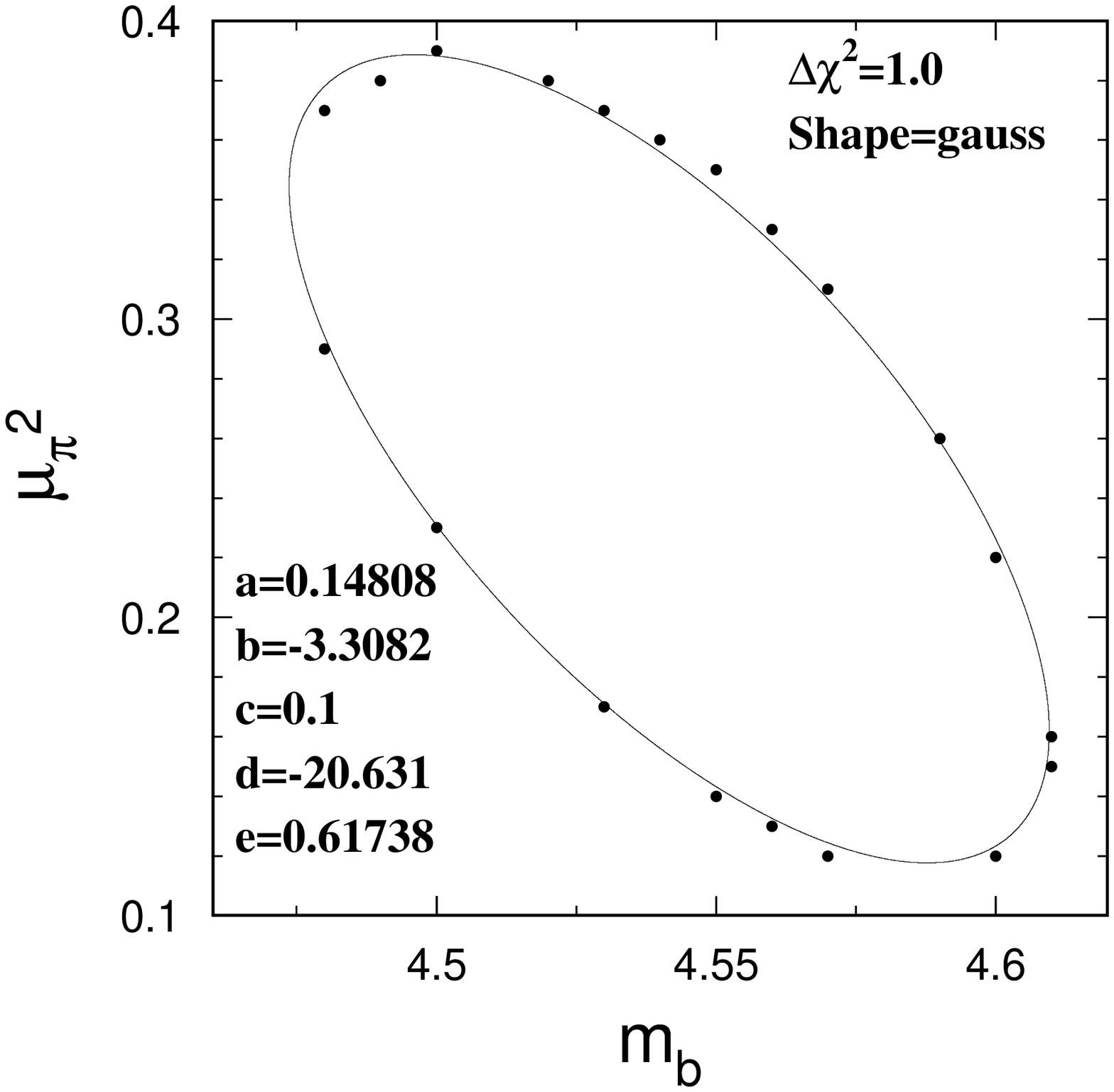} & 
    \includegraphics[width=0.33\columnwidth]{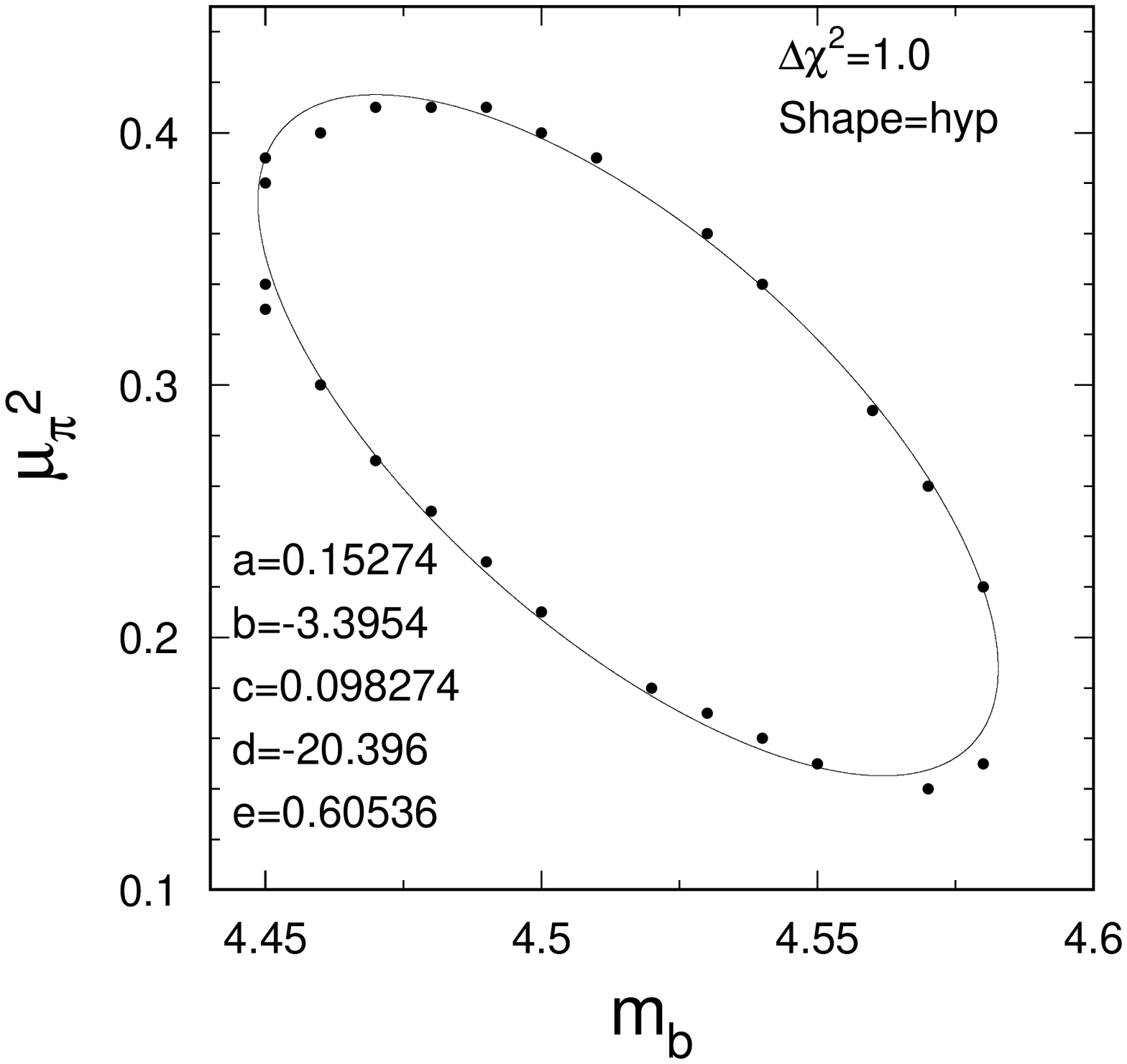} \\
    (a) exponential & (b) gaussian & (c) hyperbolic \\
  \end{tabular}
\end{center}
\caption[The minimum $\chi^2$ fits]{Top: Minimum $\chi^2$ fits of MC simulated
spectra to the raw data for each shape function model. \label{fig:bestfcont}
Bottom: The fitted $\Delta\chi^2=1$ contours for each shape function model.}
\end{figure}

\begin{figure}[hbpt!]
  \begin{center}
    \includegraphics[width=0.45\columnwidth]{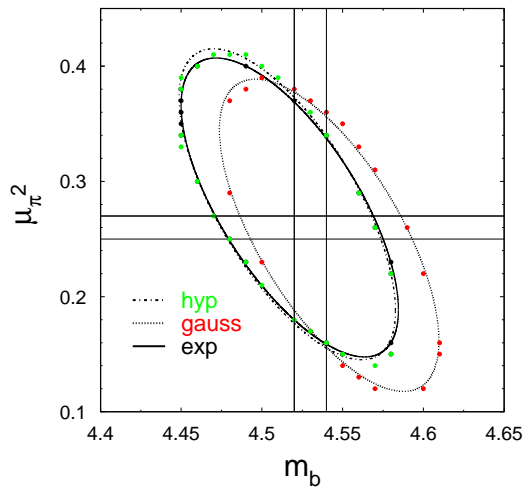}
  \end{center} \caption[ilija]{ Comparison of the fitted
  $\Delta\chi^2=1$ contours for all shape function models.  The
  contours for the exponential, gaussian and hyperbolic model are
  shown by the solid, dotted and dash-dotted curves,
  respectively. The vertical and horizontal lines mark the central values of the three fits.\label{fig:allthree}}
\end{figure}

The best fit parameters are given in \Tab{tab:besttab}. The parameter
values are found to be consistent across all three shape function
forms. The minimum $\chi^2$ fit for each shape function model is
displayed in \Fig{fig:bestfcont}. The fits to the contour with
$\Delta\chi^2=1$ points are shown in \Fig{fig:bestfcont} and
\ref{fig:allthree}. The imposed shape function form acts to correlate
\mbSF\ and \mupidSF.
\begin{table}[hbpt]
  \begin{center}
  \begin{tabular}{lccc} \hline\hline                      
    Shape   &~~$\chi^2_\mathrm{min}$~~ & ~~~~$\mbSF$~~~~& ~~~~$\mupidSF$~~~~  \\ 
     &  & $\mathrm{GeV}/c^2$ & $\mathrm{GeV}^2/c^2$  \\ \hline
    exponential & 4.32 & 4.52  & 0.27  \\
    gaussian    & 3.78 & 4.54  & 0.25  \\
    hyperbolic  & 4.41 & 4.52  & 0.27  \\ \hline\hline
  \end{tabular} 
  \end{center}
  \caption{The best fit shape function parameter values.\label{tab:besttab}}
\end{table}


\section{Summary}
The $b$-quark leading shape function parameters in the shape function
scheme, \mbSF\ and \mupidSF, were determined from fits of Monte Carlo
simulated spectra, generated by the prescription in
Ref.~\cite{generator}, to the raw~\footnote{ Raw refers to the
spectrum as measured after the application of analysis cuts.} Belle
measured \bsg\ photon energy spectrum. Three models for the form of
the leading shape function were used; exponential, gaussian and
hyperbolic, while the default model from Ref.~\cite{generator} was
used for the subleading shape function, where the reference scale is
chosen to be $1.5 \gev$.
Best fit parameters are:
$(\mbSF,\mupidSF)_{\mathrm{exp}}=(4.52,0.27)$,
$(\mbSF,\mupidSF)_{\mathrm{gauss}}=(4.54,0.25)$, and
$(\mbSF,\mupidSF)_{\mathrm{hyp}}=(4.52,0.27)$, where \mbSF\ and
\mupidSF\ are measured in units of $\mathrm{GeV}/c^2$ and
$\mathrm{GeV}^2/c^2$ respectively. We also determined the $\Delta
\chi^2 = 1$ contours in the $(\mbSF,\mupidSF)$ parameter space for
each of the assumed models.
    
\section*{ACKNOWLEDGMENTS}
We would like to thank all Belle collaborators, in particular Patrick
Koppenburg. We acknowledge support from the Ministry of Education,
Culture, Sports, Science, and Technology of Japan and the Japan
Society for the Promotion of Science; the Ministry of Higher
Education, Science and Technology of the Republic of Slovenia.

We are grateful to B. Lange, M. Neubert and G. Paz for providing us
with their theoretical computations implemented in an inclusive
generator. We would specially like to thank M. Neubert for valuable
discussions and suggestions.


\begin{thebibliography}{99}
\bibitem{Gibbons}  
  L.~Gibbons  [CLEO Collaboration], AIP Conf.\ Proc.\  {\bf 722}, 156 (2004).
\bibitem{limonoz}
  A.~Limosani and T.~Nozaki [Heavy Flavor Averaging Group], arXiv:hep-ex/0407052.
\bibitem{KN}
   A.L.  Kagan and M. Neubert, Eur. Phys. J. \textbf{C7} 5 (1999).
\bibitem{BLNP}
  S.W. Bosch, B.O. Lange, M. Neubert and G. Paz, Nucl. Phys. Nucl.\ Phys.\ B {\bf 699}, 335 (2004);
  M. Neubert, Eur. Phys. J. C (in print) [arXiv:hep-ph/0408179] ;
  S.W. Bosch, M. Neubert and G. Paz, JHEP {\bf 0411}, 073 (2004);
  M. Neubert, arXiv:hep-ph/0411027;
  M. Neubert, arXiv:hep-ph/0412241.
\bibitem{generator} 
  B.O.~Lange, M.~Neubert and G.~Paz, hep-ph/0504071 and private communication with M.~Neubert.
\bibitem{endpoint}
  A.~Limosani {\it et al.} [Belle Collaboration], arXiv:hep-ex/0504046.
\bibitem{fullrec}
  I.~Bizjak {\it et al.} [Belle Collaboration], arXiv:hep-ex/0505088.
\bibitem{Anderson}  
 S. Anderson, Ph.D. thesis, University of Minnesota, 2002.
\bibitem{Koppenburg}
  P.~Koppenburg {\it et al.}  [Belle Collaboration], Phys.\ Rev.\ Lett.\  {\bf 93}, 061803 (2004).
\bibitem{lange}  
 We thank B. Lange for noticing us this fact.
\bibitem{Fac}  
 We thank R. Faccini for suggesting such a function.
\end{thebibliography}
\end{document}